\documentclass{amsart}
\usepackage{natbib}

\usepackage{a4wide}
\usepackage[dmyyyy]{datetime}

\usepackage{graphicx}
\usepackage{latexsym}
\usepackage{amssymb}
\usepackage{amsmath}
\usepackage{comment}

\usepackage[colorinlistoftodos]{todonotes}  % for draft

\usepackage{pdfsync}

\usepackage[font=small]{caption}

\newcommand\membership{\textit{memb}\xspace}
\newcommand\forward{\textit{fwd}\xspace}
\newcommand\backward{\textit{bwd}\xspace}

\newcommand\maxlen{\textit{maxlen}}
\newcommand\shorter{\textit{shorter}}
\newcommand\longer{\textit{longer}}
\newcommand\lastchar{\textit{lastchar}}

\newcommand\minsert{\textit{insert}\xspace}
\newcommand\mdelete{\textit{delete}\xspace}

\usepackage[ruled,linesnumbered]{algorithm2e}
\usepackage{hyperref}

\newcommand{\kmer}{$k$-mer\xspace}
\newcommand{\kmers}{$k$-mers\xspace}

\newcommand\subprop{spectrum-like-property\xspace}
\newcommand\bigoh{{\mathcal O}\xspace}

\definecolor{deepjunglegreen}{rgb}{0.0, 0.29, 0.29}

\begin{document}

\title{Data structures to represent a set of $k$-long DNA sequences} 
\author{Rayan Chikhi}
\address{Rayan Chikhi, Center of Bioinformatics and Biostatistics and Integrative Biology - USR 3756, Institut Pasteur and CNRS, 25-28 rue du Docteur Roux, 75015 Paris, France}
\email{rayan.chikhi@pasteur.fr}
\author{Jan Holub}
\address{Jan Holub, Department of Theoretical Computer Science, Faculty of Information Technology, Czech Technical University in Prague, Th{\'a}kurova 2700/9, 160\,00, Prague~6, Czech Republic}
\email{Jan.Holub@fit.cvut.cz}
\author{Paul Medvedev}
\address{Paul Medvedev, Department of Computer Science and Engineering and Department of Biochemistry and Molecular Biology, 506B Wartik Lab, University Park, The Pennsylvania State University, PA 16802, USA}
\thanks{This research has started during J.~Holub's research stay at the Pennsylvania State University supported by the Fulbright Visiting Scholar Program and it was finished with the support of the OP VVV MEYS funded project CZ.02.1.01/0.0/0.0/16\_019/0000765 ``Research Center for Informatics''.  This work was partially supported by NSF awards DBI-1356529, CCF-551439057, IIS-1453527, and IIS-1421908 to PM. Research reported in this publication was supported by the National Institute Of General Medical Sciences of the National Institutes of Health under Award Number R01GM130691, and the INCEPTION project (PIA/ANR-16-CONV-0005). The content is solely the responsibility of the authors and does not necessarily represent the official views of the National Institutes of Health. }
\email{pzm11@psu.edu}

\maketitle

\begin{abstract}
	The analysis of biological sequencing data has been one of the biggest applications of string algorithms.
	The approaches used in many such applications are based on the analysis of \kmers, which are short fixed-length strings present in a dataset.
	While these approaches are rather diverse, storing and querying a \kmer set has emerged as a shared underlying component.
	A set of \kmers has unique features and applications that, over the last ten years,
	have resulted in many specialized approaches for its representation.
	In this survey, we give a unified presentation and comparison of the data structures that have been proposed
	to store and query a \kmer set.
	We hope this survey will serve as a resource for researchers in the field as well as make the area more accessible to researchers outside the field.
	%We hope this survey will not only serve as a resource for researchers in the field but also make the area more accessible to outsiders.
\end{abstract}

\section{Introduction} \label{sec@intro}
String algorithms have found some of their biggest applications in modern analysis of sequencing data.
Sequencing is a type of technology that takes a biological sample of DNA or RNA and extracts many {\em reads} from it.
Each read is a short substring (e.g. anywhere between 50 characters and several thousands of characters, or more)
of the original sample, subject to errors.
Analysis of sequencing data relies on string matching with these reads, 
and many popular methods are based on first identifying short, fixed-length substrings of the reads.
These are called \kmers, where $k$ refers to the length of the substring; equivalently, some 
papers use the term $q$-gram instead of \kmer.
Such \kmer-based methods have become more popular in the last ten years 
due to their inherent scalability and simplicity.
They have been applied across a wide spectrum of biological domains, e.g.
genome and transcriptome assembly, transcript expression quantification, metagenomic classification, structural variation detection, and genotyping.
While the algorithms working with \kmers are rather diverse, storing and querying a set of \kmers
has emerged as a shared underlying component. 
Because of the massive size of these sets, minimizing their storage requirements and query times is becoming its own area of research.

In this survey, we describe published data structures for indexing a set of \kmers such that set membership can be checked either directly
or by attempting to extend elements already in the set (called navigational queries, to be defined in Section~\ref{sec:operations}).
We evaluate the data structures based on their theoretical time for membership and navigational queries, space and time for construction, and time for insertion or deletion.
We also describe known lower bounds on the space usage of such data structures and 
various extensions that go beyond membership and navigational queries. 
We do not describe the various applications of \kmer sets to biological problems, i.e. strategies for constructing the \kmer set from the biological data (e.g. sampling, error detection, etc.) or algorithms that use \kmer set data structures to solve some problem (e.g. assembly, genotyping, etc\ldots).
An example of an application we do not specifically discuss is the use of a \kmer set as an index, 
e.g. when a \kmer is used to retrieve a position in a reference genome.

Since these data structures are often developed in an applied context and published outside the theoretical computer science community, 
they do not consistently contain thorough mathematical analysis or even problem statements.
There is the additional problem of inconsistent definitions and terminology.
In this survey, we attempt to unify them under a common set of query operations, categorize them, and draw connections between them.
We present a combination of: 1) non-specialized data structures (i.e. hash tables) that have been applied to \kmer sets as is,
2) non-specialized data structures that have been adapted for their use on \kmer sets, and 
3) data structures that have been developed specifically for \kmer sets.
We give a high level overview of all categories, but we give a more detailed description for the third category.
The survey can be read with only an undergraduate-level understanding of computer science, 
though knowledge of the FM-index would lead to a deeper understanding in some places.

Let $S$ denote a set of $n$ \kmers.
Representing a set is a well-studied problem in computer science.
However, the fact that the set consists of strings, and that the strings are fixed-length, lends structure that can be exploited for efficiency.
There are other factors as well.
First, in most applications, the alphabet has constant size, denoted by $\sigma$.
Second, most applications revolve around 
sets where $n = o(\sigma^k)$;
in this survey, we refer to these as {\em sparse} sets.
Third, $n$ is typically much larger than $k$, e.g. 
$k$ is usually between $20$ and $200$, while $n$ can be in the billions.

Another unique aspect of a \kmer set is what we call the {\em \subprop}.
$S$ has the {\em \subprop} if there exists a collection ${\mathcal G}$ of long strings that ``generates'' $S$. 
By ``generates'', we mean that $S$ contains a significant portion of the \kmers of ${\mathcal G}$, and, conversely,
many of the \kmers of $S$ are either exact or ``noisy'' substrings of ${\mathcal G}$.
${\mathcal G}$ is usually unknown.
For example, sequencing a metagenome sample 
($\mathcal G$ would be the set of abundant genomes in this case) generates a set of reads,
which cover most of the abundant genomes in the sample.
A computational tool would then chop the reads up into their constituent \kmers (e.g. $k=50$), 
and store these in the set $S$.
Some other examples of $\mathcal G$ are a single genome (e.g. whole genome sequencing),
a collection of transcripts (RNA-seq or Iso-Seq), or enriched genomic regions (e.g. ChIP-seq).
We introduce this property in order to informally capture an important aspect of $S$ 
in many applications that arise from sequencing. 
Our definition is necessarily imprecise, in order to capture the huge diversity in how sequencing technologies are applied and how sequencing data is used.
However, as we will show, this property is exploited by methods for representing a \kmer set and also drives the types of queries that are performed on them.

%%%%%%%%%%%%%%%%%%%%%%%%%%%%%%%%%%%%%%%%%%%%%%%%%%%%%%%%%%%%%%%%%%%%%%%%%%%%%%%%
\section{Operations}\label{sec:operations}
In this section, we describe 
a common set of operations that unifies many of the data structures for representing a set of \kmers.
%the type of operations supported by data structures representing $S$.
First, let us assume that the size of the alphabet ($\sigma$) is constant, all $\log$s are base 2, 
strings are 1-indexed, and $S$ is sparse.
The most basic operations that a data structure representing $S$ supports are its construction 
and checking whether a \kmer $x$ is in $S$ ($\membership$, which returns a boolean value).
If the data structure is {\em dynamic}, it also supports inserting a \kmer into $S$ ($\minsert$) or deleting a \kmer from $S$ ($\mdelete$).
A data structure where insertion and deletion is either not possible or would require as much time as re-construction is called {\em static}.

Recall that in the context of the \subprop, there is an underlying set of strings $\mathcal G$ that is generating the \kmers of $S$.
This implies that many \kmers in $S$ will have {\em dovetail} overlaps with each other
(i.e. the suffix of one \kmer equals to the prefix of another),
often by $k-1$ characters.
Algorithms that use $S$ in order to reconstruct $\mathcal G$ often work by starting from a \kmer
and extending it one character at a time to obtain the strings of $\mathcal G$.
This motivates having efficient support for operations that check if an extension of a \kmer exists in $S$.
A forward extension of $x$ is any \kmer $y$ such that $y[1,k-1] = x[2,k]$, and
a backward extension is any \kmer $y$ such that $y[2,k]=x[1,k-1]$
(we use the notation $x[i,j]$ to refer to the substring of $x$ starting from the $i^\text{th}$ character up to and including the $j^\text{th}$ character). 
Formally, given $x \in S$ and a character $a$,
the $\forward(x,a)$ operation  returns true if $x[2,k] \cdot a$ is in $S$
(we use $\cdot$ to signify string concatenation).
Similarly, the $\backward(x,a)$ operation checks whether $a \cdot x[1,k-1]$ is in $S$.
We refer to \forward and \backward operations as navigation operations.

We assume that a data structure maintains some kind of internal state corresponding to the last queried \kmer
i.e. a $\membership(x)$ query would leave the data structure in a state corresponding to $x$, 
a $\forward(x,a)$ query would leave the state corresponding to $x[2,k]\cdot a$, etc.
For example, for a hash table, the internal state after a $\membership(x)$ query would correspond to the hash value of $x$ and to the memory location of $x$'s slot;
in the case of an FM-index or a similar data structure, the internal state corresponds to an interval representing $x$.

We also assume that prior to a call to $\forward(x,a)$ or $\backward(x,a)$, 
the data structure is in a state corresponding to $x$.
In this way, $\forward(x,a)$ and $\backward(x,a)$ are different from $\membership(x[2,k]\cdot a)$ and $\membership(a \cdot x[1,k-1])$, respectively.
For example, it would be invalid to execute $\forward(ACG, T)$ after executing $\membership(CCC)$ because the \membership operation would leave the data structure in a state corresponding to $CCC$ 
and executing $\forward$ requires it to be in a state corresponding to $ACG$.
For data structures that do not support \forward or \backward explicitly or do not maintain an internal state, 
there is always the default implementation using the corresponding membership query.

In the following, we will first summarize some basic data structures for the above problem (Section \ref{sec:basic}). 
In Section~\ref{sec:dbg}, we will make the connection to de Bruijn graphs and present data structures that aim for fast \forward and \backward queries.
In Section~\ref{sec:navigational}, we present special type of data structures where \membership queries are very expensive or impossible, but navigational queries are cheap.
We summarize the query, construction, and modification time and space complexities of the key data structures in 
Tables~\ref{table_queries} and~\ref{table_modifications}.
In the Appendix, we show how these complexities are derived for the cases when it is not explicit in the original papers.
We then continue to other aspects.
In Section~\ref{sec:lower}, we describe the known space lower bounds for storing a set of \kmers.
Finally, in Section~\ref{sec:variations}, we describe various variations on and extensions of the data structures presented in Sections~\ref{sec:basic}--\ref{sec:navigational}.

We note that the definition of $S$ as a set implies that there is no count information associated with a \kmer in $S$.
However, some of the data structures we will present also support maintaining count information with each \kmer.
Rather than present how this is done together with each data structure that supports it, 
we have a separate section (Section~\ref{sec:counts}) dedicated to how the presented data structures can be adapted to store count information.

\section{Basic approaches}\label{sec:basic}
Perhaps the most basic static representation that is used in practice is 
a lexicographically {\bf sorted list} of \kmers. The construction time is $\bigoh(nk)$ using any linear time string sort algorithm
and the space needed to store the list is $\Theta(nk)$. 
A membership query is executed as a binary search in time $\bigoh(k\log n)$.
This representation is both space- and time-inefficient,
as it is dominated by other approaches we will discuss (e.g. unitig-based approaches or BOSS).
But it can be used by someone with very limited computer science background, 
making it still relevant.

Sorted lists can be partitioned in order to speed-up queries.
In this approach, taken by~\citet{kraken}, the \kmers are partitioned according to a minimizer function.
For a given $\ell < k$, 
an $\ell$-minimizer of a \kmer $x$ is the smallest 
(according to some given permutation function) 
$\ell$-mer substring of $x$~\citep{schleimer,roberts}. 
A minimizer function is a function that maps a \kmer to its minimizer, or, equivalently, to an integer in 
$\{1, \dots, \sigma^\ell\}$.
In the partitioned sorted list approach, 
the \kmers within each partition are stored in a separate sorted list,
and a separate direct-access table maps each partition to the location of the stored list.
In order for this table to fit into memory, $\ell$ should be small (e.g. $\ell=13$ for $\sigma=4$).
This approach can work well to speed up queries when there are not many \kmers in each partition. 
However, the space used is still $\Theta(nk)$, which is inefficient compared to more recent methods we will present.

Two traditional types of data structures to represent sets of elements are binary search trees and hash tables.
A binary search tree and its variants require $\bigoh(\log n)$ time for membership queries and are in most aspects worst than a string trie~\citep{makinen2015genome}.
To the best of our knowledge, binary search trees have not been used for directly indexing \kmers.
In a {\bf hash table}, the amortized time for a membership query, insertion, deletion, \forward and \backward is equivalent to the time for hashing a \kmer~\citep{clrs}.
Hashing a \kmer generally requires $\bigoh(k)$ time, but
one can also use rolling hash functions.
In a rolling hash function~\citep{lemire2010recursive}, if we know the hash value for a \kmer $x$, 
we can compute the hash value of any forward or backward extension of $x$ in $\bigoh(1)$ time.
Using a rolling hash function can therefore improve the \forward/\backward query time to $\bigoh(1)$.
These fast query and modification times and the availability of efficient and easy-to-use hash table libraries in most popular programming languages 
make hash tables popular in some applications.
However, a hash table requires $\Theta(nk)$ space, which is prohibitive for large applications due to the $k$ factor.

{\bf Conway and Bromage}~\citep{CB11} were one of the first to consider more compact representations of a \kmer set.
$S$ can be thought of as a binary bitvector of length $\sigma^k$,
where each \kmer corresponds to a position in the bitvector and the value of the bit reflects whether the \kmer is present in $S$.
Since $S$ is sparse, storing the bitvector wastes a lot of space.
The field of compact data-structures~\citep{compactbook} concerns exactly with how to store such bitvectors space-efficiently.
In this case, a sparse bitmap representation 
\citep{okanohara2007practical}
based on Elias-Fano coding~\citep{Eli74} 
can be used to store the bitvector; 
then, the \membership operation becomes a pair of rank operations
(i.e. finding the number ones in a prefix of a bitvector)
on the compressed bitvector.
However, if $S$ is {\em exponentially sparse}
(i.e. $\exists \epsilon > 0 \text{ such that } n = \bigoh(\sigma^{k(1-\epsilon)})$),
the space needed is $\Omega(nk)$.

\subsection{Approximate membership query data structures}\label{sec:approximate}
An approximate membership query data structure is a type of probabilistic data structure that represents a set in a space-efficient manner in exchange 
for allowing membership queries to occasionally return false positives (no false negatives are allowed though).
A false positive occurs when $x\notin S$ but $\membership(x)$ returns true.
These data structures are applicable whenever space savings outweigh the drawback of allowing some false positives or 
when the effect of false positives can be mitigated using other methods.
Note that approximate membership queries are not related to the type of queries which ask 
whether $S$ contains a \kmer with some bounded number of mismatches (e.g. one substitution) to the query \kmer.

{\bf Bloom filters}~\citep{BloomFilter} (abbreviated BF) are a classical example of an approximate membership data structure
that has found widespread use in representing a \kmer set
(see~\citet{broder2004network} for a definition and analysis of Bloom filters).
Some of the earliest applications were by \citet{facs,shi2010} 
BFs applied to \kmers support \minsert, \membership, \forward, and \backward operations in the time it takes to hash a \kmer 
(usually $\Theta(k)$, except for rolling hash functions)
and take $\bigoh(n)$ space.
A BF does not support $\mdelete(x)$, though there are variants of BFs that make trade-offs to support it in $\Theta(k)$ time
(e.g. counting BFs~\citep{countingbf} and spectral BFs~\citep{spectralbf}. 
Further time-space tradeoffs can be achieved by compressing a BF using RRR~\citep{rrr} enconding~\citep{compressedbf}.
See~\citet{tarkoma2011theory} for a survey of BF variations and the trade-offs they offer.

\citet{pellow2016} developed several modifications of a Bloom filter, specifically for a \kmer set.
They take advantage of the spectrum-like property to either reduce the false positive rate or decrease the space usage.
The general idea is that when $S$ has the spectrum-like property, most of its \kmers will have some backward and forward extension present in $S$. 
The (hopefully small amount of) \kmers for which this is not true are maintained in a separate hash table. 
For the rest, in order to determine whether a \kmer $x$ is in $S$, 
they make sure the BF contains not only $x$ but also at least one forward and one backward extension.
Using similar ideas, they give other versions of a BF for when $S$ is the spectrum of a read set or of one string. 
In another paper, \citet{chu2018improving} developed what they called a multi-index BF, which similarly takes advantage of the spectrum-like property (details omitted).

Bloom filters are popular because they reduce the space usage to $\bigoh(n)$ while maintaining $\bigoh(k)$ membership query time.
BFs and their variants are also valuable for their simplicity and flexibility.
However, operations on Bloom filters generally require access to distant parts of the data structure, and
therefore do not scale well when they do not fit into RAM.
Here, we highlight some more advanced approximate membership data structures offer better performance and have been applied to \kmers sets.
There is the quotient filter~\citep{qf} and the counting quotient filter~\citep{cmq}, 
which have been applied to storing a \kmer set in~\citet{squeakr} and~\citet{mantis}.
There is also the quasi-dictionary \citep{marchet2018resource}
and $\ell$-Othello~\citep{liu2018novel}, both generally applicable to any set of elements but applied to a \kmer set by the authors.
Cuckoo filters~\citep{cuckoo} are another approximate membership data structure that has been applied to \kmers~\citep{zentgraf2020cost}.

\subsection{String-based indices}\label{sec:string}
There is a rich literature of string-based indices~\citep{makinen2015genome}, some of which can be modified to store and query a \kmer set.
One of the most popular string-based indices to be applied to bioinformatics is the
{\bf FM-index}\footnote{We note that the FM-index and its variants are also sometimes referred to as a BWT-indices, since they are based on the Burrows-Wheeler Transform (BWT).}\citep{FM2000}.
It can be defined and constructed for a set of strings,
using the Extended Burrows-Wheeler Transform~\citep{mantaci2005extension}.
A scalable version has been implemented in the BEETL software~\citep{bauer2013lightweight}.
This can in principle be applied to $S$ (by treating every \kmer in $S$ as a separate string), 
though we are unaware of such an application in practice.
In theory, it results in $\bigoh(nk)$ construction time and $\bigoh(k)$ \membership query time~\citep{bauer2013lightweight}.
A naive implementation of \forward and \backward operations in this setting would require a new \membership query;
however, we hypothesize that a more sophisticated approach, using bidirectional indices, may improve the run-time 
(this however does not appear in the literature and is not proven).
However, the FM-index is not usually directly used for storing \kmers;
	rather, it is either used in combination with other strategies (e.g. DBGFM and deGSM,
	which we will describe in Section~\ref{sec:unitig}) 
	or in a form specifically adapted to \kmer queries (i.e. the BOSS structure, which we will describe in Section~\ref{sec:node}).

Another popular string-based index is the trie data structure and its variations. 
A trie is a tree-based index known for its fast query time, with strings labeling nodes and/or edges (see~\citet{makinen2015genome} for details).
Tries have been adopted to the \kmer set setting in a data structure called 
the {\bf Bloom filter trie}~\citep{holley2016bloom}.
It combines the elements of Bloom filters and burst tries~\citep{bursttrie}.
Conceptually, a small parameter $\ell < k$ is chosen and 
all the \kmers are split into $k/\ell$ equal-length parts.
The $i$-th part is then stored within a node at the $i$-th level of the trie.
Bloom filters are used within nodes to quickly filter out true negatives when querying the membership of a $k$-mer part.
The Bloom filter trie offers fast \membership time ($\bigoh(k)$) but requires $\bigoh(nk)$ space.

\section{De Bruijn graphs}\label{sec:dbg}

\begin{figure}[t]
\centering
\includegraphics[scale=0.5]{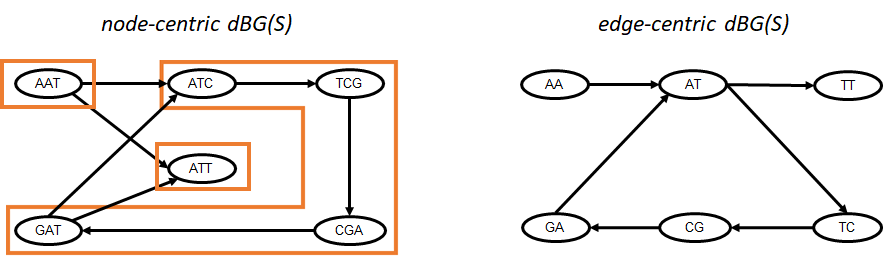}
\caption{
An example of the node-centric de Bruijn graph (left) and the edge-centric one (right).
Both graphs are built for $S = \{AAT,ATC,TCG,CGA,GAT,ATT\}$.
There are three maximal unitigs in the node-centric graph,
highlighted in the figure with orange rectangles.
The spellings of the three maximal unitigs are $AAT$, $ATCGAT$, and $ATT$.
}
\label{fig:dbg}
\end{figure}

A de Bruijn graph provides a useful way to think about a \kmer set that has the \subprop and for which \forward and \backward operations should be supported more efficiently than membership operations. %in $\bigoh(1)$ time.
A de Bruijn graph (dBG) is directed graph built from a set of \kmers $S$. In the {\em node-centric} dBG, the node set is given by $S$ and there is an edge from $x$ to $y$ iff the last $k-1$ characters of $x$ are equal to the first $k-1$ characters of $y$. In a {\em edge-centric} dBG, the node set is given by the set of $(k-1)$-mers present in $S$, and, for every $x\in S$, there is an edge from $x[1,k-1]$ to $x[2, k]$. 
In other words, the \kmers of $S$ are nodes in the node-centric dBG and edges in the edge-centric dBG.
Figure~\ref{fig:dbg} shows an example.
The graphs represent equivalent information.
Technically, the node-centric dBG of $S$ is a line graph~\citep{digraphs} of the edge-centric dBG of $S$, and without loss-of-generality,
we mostly focus our discussion on node-centric dBGs.

The concept of a de Bruijn graph in bioinformatics
is originally borrowed from combinatorics, 
where it is used to denote the node-centric dBG (in the sense we define here) of the full \kmer set, 
i.e. a set of all $\sigma^k$ \kmers.
It found its initial application in bioinformatics in genome assembly algorithms~\citep{simpson2015theory}.
We do not discuss this application here, but rather we discuss its relationship to the representation of a \kmer set.

The dBG is a mathematical object constructed from $S$ that explicitly captures the overlaps between the \kmers of $S$.
Since this information is already implicitly present in $S$, the dBG contains the same underlying information as $S$.
However, the graph formalism gives us a way to apply graph-theoretic concepts, such as walks or connected components, to a \kmer set. 
In theory, all these concepts could be stated in terms of $S$ directly without the use of the dBG. For example a simple path in the node-centric dBG could be defined as an ordered subset of $S$ such that every consecutive pair of \kmers $x$ and $y$ obey $x[2,k] = y[1,k-1]$. However, using the formalism of de Bruijn graphs makes the use of graph-theoretic concepts simpler and more immediate.

Just like $S$ is a mathematical object that can be represented by various data structures, so is the dBG.
In this sense, the term dBG can have a fuzzy meaning when it is used to refer to not just the mathematical object
but to the data structure representing it. 
Generally, though, when a data structure is said to represent the dBG (as opposed to $S$),
it is meant that edge queries can be answered efficiently.
When projected onto the operations we consider in this paper, 
in- and out-edge queries are equivalent to \backward and \forward queries, respectively. 
In particular, a query to check if $x$ has an outgoing edge to $y$
is equivalent to the $\forward(x, y[k])$ operation,
while $\forward(x,a)$ is equivalent to checking if $x$ has an outgoing edge to $x[2,k]\cdot a$.

\subsection{Node- or edge-based representations}\label{sec:node}
The simplest data structures that represent graphs are the incidence matrix and the adjacency list. 
The incidence matrix representation requires $\Theta(n^2)$ space and is rarely used for dBGs
(the inefficiency can also be explained by the fact 
the incidence matrix is not intended for sparse graphs, but
the dBG is sparse because its nodes have constant in- and out-degrees of at most $\sigma$).
A {\bf hash table adjacency list} representation is possible using a hash table that stores, 
for each node, 
$2\sigma$ bits to signify which incident edges exist in the graph.
Concretely, each node $u$ potentially has $2\sigma$ outgoing edges, 
corresponding to the possible $\sigma$ forward extensions and $\sigma$ backward extensions.
Thus, we can use one bit for each of the $2\sigma$ possible edges to indicate their presence/absence.
The navigational operations still require the time needed to hash a \kmer because
the hash value for the extension needs to be calculated in order to change the ``internal state'' of the hash table to the extension.
However, checking which extensions exist can be done in constant time.
While this representation requires $\Theta(nk)$ space, 
its ease of implementation makes it a popular choice for smaller $n$ or $k$.

The special structure of dBGs (relative no arbitrary graphs) has been exploited to create a more space-efficient representation called {\bf BOSS}
(the name comes from the initials of the inventors~\citep{BOSS2012}).
BOSS represents the edge-centric dBG as a list of the edges' extension characters (i.e. for each edge $x$, 
the character $x[k]$), 
sorted by the concatenation of the 
reverse of the source node label and the 
extension character (i.e. $x[k-1]\cdot x[k-2] \cdots x[1] \cdot x[k]$).
The details of the query algorithm are too involved to present here, and we refer the reader to either the original paper or to~\citet{makinen2015genome}.
BOSS builds upon the XBW-transform \citep{FLMM2009} representation of trees, which itself is an extension of the FM-index \citep{FM2000} for strings.  
BOSS further modified the XBW-transform to work for dBGs. 
Historically, BOSS was initially introduced such that it was computed on a single string as input~\citep{BOSS2012}; then an efficient implementation used $k$-mer-counted input (COSMO, \citet{BBGPS2015}); finally some modifications have been made to the original structure for usage in a real genome assembler \citep{li2016megahit}.

BOSS occupies $4n+o(n)$ bits of space and allows operation $\membership(u)$ in ${\mathcal O}(k)$ time,
which works like the search operation in an FM-index \citep{FM2000}. 
This assumes that there is only one source and one sink in the dBG. 
If there are more sources and sinks in the dBG but their number is negligible, 
the space becomes $5n + o(n)$ (this is due to a distinct separator character being needed, as described in~\citet{BOSS2012}). 
Otherwise, in the worst case, the space needed becomes $\Theta(nk)$~\citep{BOSS2012,BBGPS2015}.
In the version given by~\citet{li2016megahit}, the space is always $6n + \bigoh(1)$, but then membership queries 
sometimes give incorrect answers.
BOSS achieves a $\bigoh(1)$ run-time for the \forward operation, 
while \backward still runs in $\bigoh(k)$ time. 
The \backward query time can further be reduced to $\bigoh(1)$ using the method of \citet{biBOSS}, at the cost of $\bigoh(n)$ extra space.
This representation is static, but a dynamic one is also possible by sacrificing some query time~\citep{BOSS2012,biBOSS}. 
Like approximate membership data structures, BOSS achieves $\bigoh(n)$ space and $\bigoh(k)$ \membership  query time.
The main difference is that approximate data structures have false positives while BOSS only achieves the $\bigoh(n)$ space when the number of sources/sinks is small.

\subsection{Unitig-based representations}\label{sec:unitig}
A {\em unitig} in a node-centric dBG is a path over the nodes $(x_1, \ldots, x_\ell)$, with $\ell\geq 1$ such that either (1) $\ell=1$, or
(2) for all $1 < i < \ell$, the out- and in-degree of $x_i$ is 1 and the in-degree of $x_\ell$ is 1 and the out-degree of $x_1$ is 1.
A unitig is {\em maximal} if the underlying path cannot be extended by a node while maintaining the property of being a unitig.
The set of maximal unitigs in a graph is unique and forms a node decomposition of the graph (Lemma 2 in \citet{CLM2016}).
See Figure~\ref{fig:dbg} for an example of maximal unitigs.
In the literature, maximal unitigs are sometimes referred to as unipaths or as simply unitigs.
Computing the maximal unitigs can also be viewed as a task of compacting together their constituent nodes in the graph; hence this is sometimes referred to as graph compaction.

A maximal unitig $(x_1, \ldots, x_\ell)$ spells a string $t=x_1 x_2[k] \cdots x_\ell[k]$ with the property
that a \kmer $x$ is a substring of $t$ iff $x \in \{x_1, \ldots, x_\ell\}$.
Thus, the list of maximal unitigs is an alternate representation of the \kmers in $S$ 
in the sense that $x \in S$ if and only if $x$ is a substring of a maximal unitig of the dBG of $S$.
This representation reduces the amount of space  since a maximal unitig represents a set of $\ell$ \kmers using $k - 1 + \ell$ characters, while the raw set of \kmers uses $k\ell$ characters.
The number of characters taken by the list is $n + U(k - 1)$, where 
$U$ is the number of maximal unitigs.
In many bioinformatic applications, $U$ is much smaller than $n$ and this representation can greatly reduce the space. 
However, since one can always construct a set $S$ with $U=n$, this representation does not yield an improvement 
when using worst-case analysis.

Given these space savings, one can pre-compute the maximal unitigs of $S$ as an initial, lossless, compression step.
This is itself a task that builds upon other \kmer set representations.
However, there are fast and low-memory stand-alone tools for compaction such as BCALM~\citep{CLM2016} or others~\citep{pan18,guo18}; 
more generally, algorithms for compaction are often presented as part of genome assembly algorithms, which are too numerous to cite here.

In order to support efficient \membership, \forward, and \backward queries, the maximal unitigs must be appropriately indexed.
The {\bf DBGFM} data structure~\citep{CLJSM2014} builds an FM-index of the maximal unitigs in order to allow \membership queries.  
In {\bf deGSM}~\citep{guo18}, the authors similarly build a BWT (which is the major component of an FM-index) of the maximal unitigs; but, they demonstrate how this can be
done more efficiently by not explicitly constructing the strings of maximal unitigs (details omitted).
These representations allow for $\bigoh(k)$ \membership queries.
For a \kmer that is not the first or last \kmer of a maximal unitig,
there is exactly one \forward and \backward extension, and it is determined by the next character in the unitig.
For such \kmers, 
these operations can be done in very small constant time, without the need to use the FM-index.
In the case that a \kmer lies at the end of its maximal unitig, 
it may have multiple extensions and they would be at an extremity of another maximal unitig. 
In this case a new \membership query is required, though more sophisticated techniques may be possible to reduce the query times.
It should be noted that these approaches, as implemented, are static;
however, it may be possible to modify them to allow for insertion and deletion.

Another approach to index unitigs is taken by {\bf Bifrost}~\citet{bifrost}, 
using minimizers. 
Bifrost builds a hash table where the keys are all the distinct minimizers of $S$ and 
the values of the locations of those minimizers in the maximal unitigs.
The membership of a \kmer is then checked by first computing its minimizer and 
then checking all the minimizer occurrences in the unitigs for a full match. 
The index is dynamic, i.e. it intelligently recomputes the unitigs and the minimizer index 
based on a \kmer insertion or deletion.

Before presenting other unitig-based indices, we make an aside to introduce minimal perfect hash functions.
Given a static set $S$ of size $n$, a hash function is perfect if its image by $S$ has cardinality $n$, 
i.e. there are no collisions. 
Furthermore, the hash function is minimal if the image consists of integers smaller or equal to $n-1$. 
Minimal perfect hash functions (MPHF) can in theory be efficiently constructed and evaluated; 
we omit the details and refer the reader to~\citep{chd} for an example.
When applied to a \kmer set $S$, one can construct an MPHF in $\bigoh(nk)$ time and store
it in $cn$ bits of space where $c$ is a small constant (around 3)~\citep{chd,bbhash}; 
calculating the hash value of a \kmer is done in $\bigoh(k)$ time.
There exists an efficient implementation of MPHF for a \kmer set, BBHash~\citep{bbhash}, 
designed to handle sets of billions of \kmers.
The advantage of a MPHF is that one can use it to associate information with each \kmer in $S$;
this is done by creating an array of size $n$ and using the MPHF value of a \kmer as its index into the array.
Unlike a hash table, this requires $\bigoh(n)$ instead of $\bigoh(nk)$ space.
The disadvantage of a MPHF is that if it is given a \kmer $x \notin S$, 
it will still return a location associated with some arbitrary $x' \in S$. 
Thus it cannot be used to test for membership without further additions.
Furthermore, support for insertions and deletions would require a dynamic perfect hashing scheme, 
yet to the best of our knowledge the only efficient implementation for large key sets~\citep{bbhash} is static.
This limitation is inherited by the MPHF-based schemes we will describe in this paper.

The {\bf pufferfish} index~\citep{putterfish} uses a MPHF 
as an alternate to the FM-index when indexing the maximal unitigs.
The MPHF along with additional information enables mapping each \kmer to its location in the maximal unitigs.
To check for membership, a \kmer $x$ is first mapped to its location;
then, $x\in S$ if and only if the \kmer at the location is equal to $x$
The pufferfish index is static, because of its reliance on the MPHF.
A similar approach is the {\bf BLight} index~\cite{blight}. 
It also uses an MPHF to map \kmers to locations in unitigs,
though it does it in a somewhat different way
(we omit the details here).

\citet{brindathesis,spss,simplitigs} recently extended the idea of unitig-based representations to {\bf spectrum-preserving string set} representations 
(alternatively, these are referred to as {\bf simplitigs}).
They observed that what makes unitigs useful as a representation is that they
contain the exact same \kmers as $S$, without any duplicates. 
They defined a spectrum-preserving string set representation as any set of strings 
that has this property
and gave a greedy algorithm to construct one.
The resulting simplitigs had a substantially lower number of characters than unitigs in practice.
To support \membership queries, simplitigs were combined with an FM-index~\citep{spss},
in the same manner that unitigs were combined with an FM-index to obtain DBGFM.

\section{Navigational data structures}\label{sec:navigational}
Many genome assembly algorithms start from a \kmer in the dBG and 
proceed to navigate the graph by following the out- and in-neighbor edges.
Membership queries are only needed to seed the start of a navigation with a \kmer.
Afterwards, only \forward and \backward queries are performed.
In this way, we can continue navigating to all the \kmers reachable from the seed.
A data structure to represent $S$ can take advantage of this access pattern in order 
to reduce its space usage, as we will see in this section.
Formally, a {\em navigational data structure} is one where \membership queries are 
either very expensive or impossible, but \forward and \backward queries are cheap
(e.g. $\bigoh(k)$).
Navigational data structures were first used by~\citet{minia} and later formalized in~\citet{CLJSM2014}. 

An {\bf MPHF in combination with a hash table adjacency list} representation of a dBG forms a natural basis 
for a navigational data structure, as follows.
This scheme was first described in the literature by \citet{fdbg} 
but was previously implemented in the SPAdes assembler~\citep{spades}.
An MPHF is first built on $S$ and then used to index a direct access table (i.e. an array).
Each entry is composed of $2\sigma$ bits indicating which incident edges exist.
For $x\in S$,
we can answer $\forward(x,a)$ and $\backward(x,a)$ queries using the table.
Given $x$'s hash value, it takes only $\bigoh(1)$ time to find out if an extension exists, but 
the queries take $\bigoh(k)$ time because a hash value has to be computed to actually navigate to the extension.
If a rolling MPHF is used, this can also take $\bigoh(1)$ time.

The {\bf list of maximal unitigs} also forms a natural basis for a navigational data structure, 
without the need of constructing any additional index to support \membership queries.
As previously described, when maximal unitigs are stored,
the $\forward$ and $\backward$ queries are trivial for most \kmers.
The exceptions occur when \forward is executed on the last \kmer in a maximal unitig 
or when \backward is executed on the first \kmer in a maximal unitig.
These extensions must be stored in a structure separate from the maximal unitigs; 
for example, the hash table adjacency list indexed by a MPHF can be used as described above.
This approach of indexing the extensions was taken by~\citet{LCBRP2016}.
When the number of maximal unitigs is significantly smaller than $n$, 
the cost of this additional structure is negligible.

Another approach to constructing a navigational data structure builds on the Bloom filter (BF).
A BF is first built to store the \kmers of $S$, but a hash table is also used to store
the \kmers that are false positives in the BF and are extensions of elements of $S$~\citep{minia}. 
This allows to avoid false positives for \forward/\backward queries by double checking the hash table.
More memory efficient approaches use a cascading Bloom filter~\citep{cascading,jackman2016abyss}, 
which is a sequence $B_1,\ldots,B_n$ of increasingly smaller Bloom filters, 
where $B_1$ is an initial Bloom filter that stores $S$ and $B_i$ ($i>1$) 
stores the \kmers that are false positives of $B_{i-1}$.
BF-based navigational data structures support exact \forward/\backward queries in $\bigoh(k)$ time 
(or $\bigoh(1)$ with a rolling hash); 
as a bonus, they can also support approximate \membership queries (they do not support \minsert operations).
In this sense, they can be viewed as a compromise between navigational and normal data structures 
that trades exact membership of non-extension \kmers for better space-efficiency.
Alternatively, they can be viewed as an augmentation of the simple Bloom filter representation to guarantee that at least the navigational queries are exact.

\citet{fdbg} proposed a mechanism to transform their navigational data structure (described earlier in this section)
into a membership data structure.
They give both a static and dynamic version; we present the static one here.
They first find a forest of node-disjoint rooted trees that is a node-covering subgraph of the dBG.
Each tree has bounded height (between $2k$ and $6k$, or less in case of a small connected component). 
They build an MPHF of $S$ and use it to store the adjacency list of the dBG, as described above.
They also use it to record, for each \kmer, 
whether it is a root in the forest and in case it is not, 
a number between 0 and $2|\Sigma|$ to represent which 
navigational query will lead to its parent.
A dictionary is used to store the node sequences of \kmers associated with each root.
Apart from these, no other node sequence is stored. 
The tree structure requires an additional $cn$ bits to store, where $c$ is implementation-dependent,
and supports membership queries in $\bigoh(k)$ time.
It is assumed that the space to store the root \kmers is a lower-order term of the whole structure, 
which is the case except when the graph consists of many small connected components.

To check for membership of a \kmer $x$, 
we start with the node $x'$ which MPHF identifies as corresponding with $x$.
We use the stored navigation instructions to follow $x'$ up to its root (using at most $6k$ queries).
If a tree root cannot be reached after $6k$ steps, 
or if any of the navigational instructions violate the information in the MPHF adjacency list, 
then we can conclude that $x' \ne x$ and hence $x \notin S$.
If a tree root is reached within $6k$ steps, $x\in S$ if and only if the sequence of the root 
(computed dynamically from traveling up the tree) is equal to the stored \kmer associated with the root.

%%%%%%%%%%%%%%%%%%%%%%%%%%%%%%%%%%%%%%%%%%%%%%%%%%%%%%%%%%%%%%%%%%%%%%%%%%%%%%%%
\begin{table}
\footnotesize
%\scalebox{0.9}{%
\setlength{\tabcolsep}{8pt}
%\rotatebox{180}{
%\begin{minipage}{\textheight}
\centering
\begin{tabular}{p{30mm}|l|l|l|}
data structure & \membership & $\forward$ &  $\backward$   \\
\hline
sorted list & %%%%%%%%%%%%%%
$k \log n$ & 
%$(k - |u|)nk|u|\log n$ &
$^\text{a}$$k \log n$ & 
$^\text{a}$$k \log n$  
\\ %%%
hash table adj. list &  %%%%%%%%%%%%%%
$k$ &
$^\text{b}1$ or $k$  &
$^\text{b}1$ or $k$ 
\\ %%%
Conway and Bromage  & %%%%%%%%%%%%%%
$\max(\log \frac{\sigma^k}{n} , \frac{\log^4 n}{k\log \sigma})$ &
$^\text{a}$ $\max(\log \frac{\sigma^k}{n} , \frac{\log^4 n}{k\log \sigma})$ & 
$^\text{a}$ $\max(\log \frac{\sigma^k}{n}, \frac{\log^4 n}{k\log \sigma})$ 
\\ %%%
Bloom filter$^1$ & %%%%%%%%%%%%%%
$k$ &
$^\text{b}1$ or $k$  &
$^\text{b}1$ or $k$ 
\\ %%%
Bloom filter trie & %%%%%%%%%%%%%%
$k$ &
$^\text{a}$$k$ &
$^\text{a}$$k$ 
    % need to do O(k) operation for a forward query, as per BFT article: "the maximum number of visited vertices in t for all k-mers of succ(x, G) is 1+heightmax(t)" and "height_max(t) = k/l + 1" (as per BFT article) and we decided that l=1 
%\dflt  % " and for all k-mers of pred(x, G) is 1+|A| heightmax(t)."

\\ %%%
BOSS (static) & %%%%%%%%%%%%%%
$k$ &
$1$ &
$1$ 
\\ %%%
BOSS (dynamic) & %%%%%%%%%%%%%%
$k(1+\frac{\log n}{\log\log n})$  &
$\frac{\log n}{\log\log n}$  & 
$k(1+\frac{\log n}{\log\log n})$ 
\\ %%%
unitig-based$^2$   & %%%%%%%%%%%%%%
$k$ &
$^\text{c}1$ or $k$ &
$^\text{c}1$ or $k$ 
\\ %%%
Belazzougui et al$^3$   & %%%%%%%%%%%%%% https://arxiv.org/pdf/1607.04909.pdf, right below Theorem 1
$k$ &  %" time to check whether a node is in G to O(k)."
$^\text{a}1$ & % "listing the edges incident to a node we are visiting takes O(σ) time". . . . BUT, we need to do more than list, we need to change states.
$^\text{a}1$  % same
\\ %%%
\hline
\end{tabular}
\caption{Query Complexities. Big O notation is implied for all the complexities, but the $\bigoh$ symbol is omitted from the table for clarity. \\
$^\text{a}$there is no specialized navigational query so the time is the same as for \membership.\\
$^\text{b}\bigoh(1)$ occurs if a rolling hash function is used, otherwise there is no specialized navigational query. \\
$^\text{c}$For DBGFM and deGSM, $\bigoh(1)$ holds if the extension lies on the same unitig;
for BLight, it holds if the extension lies on the same super-\kmer; for pufferfish, it holds if a rolling MPHF is used. \\
$^1$the Bloom filter is non-exact and may return false positives. \\
$^2$This includes DBGFM~\citep{CLJSM2014}, deGSM~\citep{guo18}, pufferfish~\citep{putterfish}, and BLight~\citep{blight}.\\
$^3$This includes both the static and dynamic version presented in~\citet{fdbg}. But,
the dynamic version may, with low probability, give incorrect query answers.
}
\label{table_queries}
%\end{minipage}
%}
\end{table}

\begin{table}%[htb]
    \setlength{\tabcolsep}{5pt}
%    \rotatebox{180}{
%    \begin{minipage}{\textheight}
    \centering
    \footnotesize
	\begin{tabular}{p{40mm}|l|l|l|l|}
	&         \multicolumn{2}{c|}{construction} & \multicolumn{2}{|c|}{modification}                        \\
	\cline{2-5}
data structure                         & time & space     & $\minsert$ &  $\mdelete$      \\
\hline
  sorted list              & $\bigoh(nk)$   & $\Theta(nk)$                       & -                  &  -                \\
  hash table adj. list             & $\bigoh(nk)$          & $\Theta(nk)$                 & $\bigoh(k)$                 & $\bigoh(k)$              \\
%  min-perfect hash table               & $O(nk)$            &          & -                   & -                  \\
Conway and Bromage        & $\Omega(nk)$              & $ \Theta(n (1 + \log \frac{\sigma^k}{n}))$    &  -                &  -              \\
  Bloom filter                         & $\bigoh(nk)$                & $\bigoh(n)$        & $\bigoh(k)$                 & -               \\
  Bloom filter trie                    & $\bigoh(nk)$                & ${\mathcal O}(nk)$ & $\bigoh(k)$                 & -               \\
  BOSS (static)     & $\bigoh(nk\frac{\log n}{\log\log n})$ & $^\text{a}\bigoh(n)$ & -                   & -              \\
BOSS  (dynamic) &  $\bigoh(nk\frac{\log n}{\log\log n})$  & $^\text{a}\bigoh(n)$   &  $\bigoh(k\frac{\log n}{\log\log n})$  &   -            \\
  %variable-order BOSS~\cite{BBGPS2015}  & static  & & $n\log k+ 4n+o(n)$ bits         & -                   & -                                       \\
%  variable-order BOSS~\cite{BBGPS2015} & dynamic &                              &                     &     &              \\
 unitig-based   & $\bigoh(nk)$          & $\bigoh(n + U(k-1))$   &  - & - \\

 Belazzougui et al (static) & $\bigoh(nk)$ & $^\text{b}\bigoh(n + kC)$ &  - & - \\ % taken from Theorem 1 of https://arxiv.org/pdf/1607.04909.pdf
 Belazzougui et al (dynamic) & $\bigoh(nk)$ & $^\text{b}\bigoh(n\log\log n + kC)$ &  ${\mathcal O}(k)$  & ${\mathcal O}(k)$ \\ % taken from Theorem 2 of https://arxiv.org/pdf/1607.04909.pdf
\hline
\end{tabular}
\caption{Construction and modification time and space complexities. Construction space refers to the size of the constructed data structure, rather than to the memory used by the construction algorithm.  \\
%$m$ is the number of edges, $N$ is the size of the input string for which de Bruijn graph is constructed. ---- both replaced by n
$^\text{a}$This assumes that either the number of sources and sinks is negligible~\citep{BOSS2012,BBGPS2015},
or the membership queries are not always exact~\citep{li2016megahit}; 
otherwise, in the worst case, the space needed is $\Theta(nk)$.\\
$^\text{b}C$ is the number of connected components in the underlying undirected dBG.
}
\label{table_modifications}
%\end{minipage}
%}
\end{table}

%%%%%%%%%%%%%%%%%%%%%%%%%%%%%%%%%%%%%%%%%%%%%%%%%%%%%%%%%%%%%%%%%%%%%%%%%%%%%%%%

\section{Space Lower Bounds}\label{sec:lower}

How many bits are necessary to store $S$, in the worst case, so that membership queries can be answered (without mistakes)? 
\citet{CB11} provided an information theoretic answer, based on the fact that to store $n$ elements from a universe of size $U$ requires $\log {U \choose n}$ bits. 
In our case, we denote this lower bound by $L(n,k) = \log {\sigma^k \choose n}$ and,
using standard inequality bounds, we have:
$$
n\log(\sigma^k/n) \leq L(n,k) \leq n\log ( \sigma^k / n) + n \log e
$$
%from bound (n/k)^k \leq (n choose k) \leq (ne/k)^k, see wikipedia.
This asymptotically matches the space of Conway and Bromage's data structure (Table~\ref{table_modifications}).
The quantity $\log (\sigma^k / n)$ reflects the density of the set, and we have that
$0 \leq \log (\sigma^k / n) \leq k \log \sigma$.
If $S$ is exponentially sparse,
then $L(n,k) = \Theta (nk)$.

\citet{CLJSM2014} explored lower bounds for navigational data structures. 
Here, how many bits are necessary to store $S$, in the worst case, so that navigational 
queries can be answered (without mistakes)? 
They showed that $L_{\textrm{nav}}(n,k) = 3.24n$ bits are required to represent a navigational data structure (for $\sigma = 4$).
Note that this beats the above lower bound for membership data structures, because a navigational 
data structure cannot answer arbitrary \membership queries. 

The above are traditional worst case lower bounds, meaning that, for any representation that uses less than $L(n,k)$ 
(respectively, $L_\textrm{nav}(n,k)$)
bits for all possible sets $S$ with $n$ elements of \kmers, there will exist at least one input where the representation 
will produce a false answer to a membership (respectively, navigational) query.
However, this is of limited interest in the bioinformatics setting, where the \kmers in $S$ come from an underlying biological 
source.
For example, the family of graphs used to prove the $L_\textrm{nav}$ bound would never occur in bioinformatics practice.
As a result, the value that worst-case lower bounds bring to practical representation of a \kmer set is limited.
In fact, the static BOSS and the static Belazzougui data structures are able to beat this lower bound in practice 
by taking advantage of a de Bruijn graph that is typically highly connected.

The difficulty of finding an alternative to worst-case lower bounds is the difficulty of modeling the input distribution. 
\citet{CLJSM2014} considered the opposite end of the spectrum.
They call $S$ {\em linear} if the node-centric de Bruijn graph of $S$ is a single unitig.
They showed that the number of bits needed to represent $S$ that is linear is $L_\textrm{lin}(n,k) = 2n$.
A linear \kmer set is in some sense the best case that can occur in practice.
However, a linear \kmer set is much easier to represent than the sets arising in practice, hence 
$L_\textrm{linear}$ it is too conservative of a lower bound.

An intermediate model was also considered by \citet{CLJSM2014}, 
where $S$ is parametrized by the number of maximal unitigs in the de Bruijn graph.
They used this parameter to describe how much space their representation takes, however, they did not pursue the
interesting question of a lower bound parametrized by the number of maximal unitigs.

An alternative to traditional worst-case lower bounds or modeling the input distribution is to 
derive more instance-specific lower bounds.
Typically, a lower bound is derived as a function of the input size, but 
a more instance-specific lower bound might be a function of the degree distribution of the de Bruijn graph 
or something even more specific to the graph structure.
These types of lower bounds are extremely satisfying when they can be used to show an algorithm is instance-optimal, 
i.e. it matches the lower bound on every instance.
\citet{spss} derive such a lower bound for the number of characters in a spectrum-preserving string set representation.
Their lower bound did not match the performance of their greedy algorithm in the worst-case,
but it came very close (within a factor of 2\%) on the evaluated input.

%%%%%%%%%%%%%%%%%%%%%%%%%%%%%%%%%%%%%%%%%%%%%%%%%%%%%%%%%%%%%%%%%%%%%%%%%%%%%%%%
\section{Variations and extensions}\label{sec:variations}
There are natural variations and extensions of data structures for storing a \kmer set, which we describe in this section. 
These are not included in Tables~\ref{table_queries} and~\ref{table_modifications}
because they do not neatly fit into the framework of those Tables.

\subsection{Membership of $\ell$-mers for $\ell < k$}
A useful operation may be to check if $S$ contains a given string $u$ of length $|u| = \ell<k$.
In some data structures, like the Bloom filter trie, it is easy to find if a \kmer begins with $u$, but 
there is no specialized way to check if $u$ appears as a non-prefix in $S$.
One way to check for $u$'s membership
is to enumerate all the \kmers in $S$ and then perform an exact string matching algorithm 
in $\bigoh(nk)$ time (e.g. Knuth-Morris-Pratt, described in the textbook of ~\citet{clrs}).
Another way is to attempt all $\sigma^{k-\ell}$ possible ways to complete a \kmer from $u$.
Both these ways are prohibitively inefficient for most applications.
However, both the static BOSS and the FM-index on top of unitigs~\citep{CLJSM2014,guo18} data structures support checking
$u$'s membership in $\bigoh(\ell)$ time; dynamic BOSS also supports this, in time $\bigoh(\ell (1 + \log n / \log \log n))$.
We omit the details of these implementation here.

\subsection{Variable-order de Bruijn graphs}
The \forward and \backward operations require an overlap of $k-1$ characters in order to navigate $S$.
However, if such an overlap does not exist, then in some applications it makes sense to 
look for a shorter overlap. 
The variable-order BOSS was introduced to allow this~\citep{BBGPS2015}.
For a given $K$, it simultaneously represents all the dBGs for $k < K$, as follows.
At any given time, the variable-order BOSS maintains an intermediate state, which is a value $k < K$ and a range
of nodes (denoted as $B$) which share the same suffix of length $k$, representing a single node in the dBG for $k$.
It supports new operations \textit{shorter}() and \textit{longer}() for changing the value of $k$ (by one),
running in ${\mathcal O}(\log K)$ and ${\mathcal O}(|B|\log K)$ time, respectively. 
The \backward operation runs in the same asymptotic time as BOSS, but $\forward$ runs in $\bigoh(\log K)$ time.
A bidirectional variable order BOSS improved that \backward operation from $\bigoh(K)$ to $\bigoh(\log K)$~\citep{biBOSS}.
The \membership times are unaffected compared to BOSS.
The space complexity is $n\log K+ 4n+o(n)$ bits, adding an extra $n\log K$ bits to the space of BOSS.

\subsection{Double strandedness}

The {\em reverse complement} of a string is the string reversed and every nucleotide (i.e. character) replaced by its Watson-Crick complement.
In many applications, it is often useful to treat a \kmer and its reverse compliment as being identical.
There are two general ways in which data structures for storing a \kmer set can be adapted to achieve this.

The first way is to make all \kmers canonical. 
A \kmer is {\em canonical} if it is lexicographically no larger than its reverse complement.
To make a \kmer $x$ canonical, one replaces it by its reverse complement if $x$ is not canonical.
The elements of $S$ are made canonical prior to construction of the data structure, 
and $\membership$ queries always make the \kmer canonical first.
This approach works well in data structures that are hash-based  
(e.g. sorted list, hash table adjacency list, Conway and Bromage, Bloom filter) or the Bloom filter trie.
The space of these data structures does not increase,
but the query times increase by the $\bigoh(k)$ operations that may be needed 
to make a \kmer canonical.

For a data structure such as BOSS, using canonical \kmers is incompatible with the specialized 
\forward and \backward operations.
For such cases, there is a second way to handle reverse complements. 
Concretely, we can compute the reverse complement closure of $S$, as follows.
We first modify $S$ by checking, for every $x\in S$, if the reverse complement of $x$ is in $S$, and, if not, adding this reverse complement to $S$.
This increases the size of the data structure by up to a factor of two, but 
maintains the same time for \forward and \backward operations.

In case of unitig based representations, the unitigs themselves can be constructed on what is called a bidirected de Bruijn graph~\citep{medvedev2007computability,medvedev2018bidirected}.
A bidirected graph naturally captures the notion of double-stranded \kmer extensions in a graph-theoretic framework.
The unitigs can then be indexed using their canonical form.
We omit the details here.

%%%%%%%%%%%%%%%%%%%%%%%%%%%%%%%%%%%%%%%%%%%%%%%%%%%%%%%%%%%%%%%%%%%%%%%%%%%%%%%%
\subsection{Maintaining \kmer counts}\label{sec:counts}
In many contexts it is natural to store a positive integer count associated with each \kmer in $S$.
Alternatively, this may be viewed as storing a multi-set instead of a set.
In the same way that a set of \kmers can be thought of as a de Bruijn graph, a multi-set of \kmers can be also thought of as a weighted de Bruijn graph.

Many of the data structures discussed naturally support maintaining counts, 
including operations to increment or decrement a count.
Any of the data structures that associate some memory location with each \kmer in $S$ can be 
augmented to store counts, e.g. a hash table adjacency list representation, a BOSS, 
or a representation based on unitigs or on a spectrum-preserving string set.
More generally, if a data structure provides a method to obtain the rank of a \kmer within $S$ (e.g. Conway and Bromage),
that rank can be used as an index into an integer vector containing the counts.
For Bloom filters, there also exist variants that allocate a fixed number of bits per \kmer to store the approximate counts
(the counting Bloom filter,~\citep{countingbf}).

The downside of such representations, however, is that they are space inefficient when the distribution of count values is skewed. 
For example, in one typical situation, most \kmers will have a count of $\leq 10$, but there will be a few with a count in the thousands. 
Since these representations use a fixed number of bits to represent a count, they will waste a lot of bits for low count 
\kmers in order to support just a few \kmers with a large count.
To alleviate this, variable-length counters can be used.
\citet{CB11} proposed a tiered approach, storing higher order bits only as needed.
More recently, the counting quotient filter~\citep{cmq} was designed with variable-length counters in mind; 
it was applied to store a \kmer multi-set by the Squeakr~\citep{squeakr} and deBGR~\citep{pandey17} algorithms.

\citet[Section~9.7.2]{makinen2015genome} also present a count-aware alternative to BOSS, 
also based on the BWT and following \citet{ValimakiR13}.
In this representation, a BWT is constructed without removing duplicate \kmers, and the count of a \kmer $x$
can then be inferred by the number of entries in the BWT corresponding to $x$.
This approach avoids storing an explicit count vector, however, it requires space to represent each extra copy of a \kmer. 
This trade-off can be beneficial when the count values are skewed and most \kmers have low counts.

\subsection{Sets of \kmer sets}

A natural extension of a \kmer set is a set of $k$-mer sets, 
i.e. $\{S_1, \ldots, S_m\}$, where each $S_i$ is a \kmer set.
Sets of \kmer sets have received significant recent interest as they are used to
index large collections of sequencing datasets or genomes from a population. 
An equivalent way to think about this is a set of \kmers $S$ where each \kmer $x$ is associated with a set of genomes
(often called colors) $c(x) \subseteq \{1\ldots m\}$.
A set of colors is referred to as a {\em color class}.
%, and a color class $j$ is said to {\em represent} a \kmer $x$ if $c(x) = j$.
If the underlying set of \kmers is intended to support navigational queries, 
then a representation of $S$ is referred to as a {\em colored de Bruijn graph}~\citep{iqbal2012novo}. 
This is an extension of viewing a \kmer set as a de Bruijn graph to the case of multiple sets.

The literature has focused on two types of queries.
The first is the basic \kmer color query: given a \kmer $x$, is $x\in S$, and, if yes, what is $c(x)$?
The second is a color matching query: 
given a set of query \kmers $Q$  and a threshold  $0 < \Theta \leq 1$,
identify all colors that contain at least a fraction $\Theta$ of the \kmers in $Q$.
%other types of queries?

Proposed representations have generally fallen into two categories.
The first explicitly stores each \kmer's color class in a way that can be indexed by the \kmer.
For example, \citet{holley2016bloom} proposed storing the color class of a \kmer at its corresponding leaf in a Bloom filter trie,
while~\citet{mantis} stored the color class in the \kmer's slot of a counting quotient filter.
Alternatively, a BOSS can be used to store the \kmers  
and the colors can be stored in an auxiliary binary color matrix $C$~\citep{vari,rainbowfish}.
Here, $C[i,j] = 1$ if the $i^\text{th}$ \kmer in the BOSS ordering has a color $j$.
Instead of using a BOSS, \kmers in the color matrix can also be indexed using a minimal perfect hash function~\citep{seqothello}
or a unitig-based representation~\citep{bifrost}.

A column of the color matrix can be viewed as binary vector specifying the \kmer membership of $S_i$.
A variation of this then replaces each column using a Bloom filter representation of $S_i$~\citep{mustafa2018,bigsi,cobs}.
Thus, each row of the color matrix becomes a position in the Bloom filter, instead of a \kmer.
This results in space savings, but representation of the color class is no longer guaranteed to be correct.

The color matrix is sometimes compressed using a standard compression technique such as RRR~\citep{rrr} or Elias-Fano encoding~\citep{vari}.
Further compression can be achieved based on the idea that, in some applications, many \kmers share the same color class.
For example, \citet{holley2016bloom},~\citet{rainbowfish}, and~\citet{mantis} assign an integer code to each color class in increasing 
order of the number of \kmers that belong to it.
Thus, frequently occurring color classes are represented using less bits.
\citet{seqothello} proposed an adaptive approach to encoding color classes.
Based on how many colors a color class contains, the class is stored as either a list of the colors, 
a delta-list encoding of the colors, or as a bitvector of length $m$.
\citet{mantismst} take advantage of the fact that
adjacent \kmers in the de Bruijn graph are likely to have similar color classes;
they then store many of the color classes not as an explicit encoding
but as a difference vector to a similar color class.
Finally, an alternative way to encode the color matrix based on wavelet trees is given by \citet{mustafa2018}.

The second category of representations are based on the Bloofi~\citep{bloofi} data structure,
which is designed to exploit the fact that many $S_i$s are similar and, more generally, many color classes have similar \kmer compositions.
Here, each $S_i$ is stored in a Bloom filter and a tree is constructed with each $S_i$ as a leaf.
Each internal node represents the union of the \kmers of its descendants, also represented as a Bloom filter.
The Bloofi datastructure was adapted to the \kmer setting by~\citet{sbt}, who called it the Sequence Bloom Tree.
The color matching query can be answered by traversing the tree top-down and pruning the search at any node where less than 
$\Theta |Q|$ \kmers match.
Further improvements were made to reduce its size and query times~\citep{howde,allsome,ssbt}.
For example, \kmers that appear in all the nodes of a subtree can be marked as such to allow more pruning during queries,
and the information about such \kmers can be stored at the root, thereby saving space~\citep{allsome,ssbt}.
Using a hierarchical clustering to improve the topology of the tree also yields space savings and better query times~\citep{allsome}.
A better organization of the bitvectors was shown to reduce saturation and improve performance~\citep{howde}.

The first category of representations are designed with the basic \kmer color query in mind, though they can be adopted to answer the color matching query as well.
The second category of methods, on the other hand, are specifically designed to answer the color matching query. 
They can be viewed as aggregating \kmer information at the color level, while the first category can be viewed as aggregating color information at the \kmer level.
For a more thorough survey of this topic, please see~\citep{marchet2019data}.

%%%%%%%%%%%%%%%%%%%%%%%%%%%%%%%%%%%%%%%%%%%%%%%%%%%%%%%%%%%%%%%%%%%%%%%%%%%%%%%%
\section{Conclusion}
In this paper, we have surveyed data structures for storing a DNA \kmer set in a way that can efficiently support membership and/or navigational queries. 
This problem falls into the more general category of indexing a set of elements, which has been widely studied in computer science.
The aspects of a DNA \kmer set that make it unique are that the elements are fixed length strings over a constant sized alphabet, the set  is sparse, and $k$ is much less than $n$.
A DNA \kmer set tends to also have what we have termed the \subprop.
This property is hard to capture with mathematical precision, but it has been a major driver behind the design of specialized data structures. 
Another way that a DNA \kmer set is different from a general set 
is that queries are sometimes more constrained than arbitrary membership queries.
In particular, navigational queries start from a \kmer that is known to be in the set and ask which of its extensions are also present.

We now give a summary of the major developments in this field.
Some methods for storing a set proved to be useful right out-of-the-box, with the major examples being hash tables, Bloom filters, and sparse bitvectors.
These methods are generic, in the sense that there is nothing specific to \kmer sets about them.
Hash tables and Bloom filters, especially, gained widespread use because of their broad software availability and conceptual simplicity, respectively.
These two offered a tradeoff between query accuracy and space; concretely, Bloom filters require only $\bigoh(n)$ space but have false positives, 
while hash tables have no false positives but require $\bigoh(nk)$ space.
They both offered fast query times of $\bigoh(k)$ for membership and $\bigoh(1)$ for navigational queries (assuming rolling hash functions are used).
Beyond these, other generic methods found applicability in \kmer sets, especially approximate membership query data structures.
These offer both practical and theoretical improvements; however, describing these requires a more fine-grained analysis than we are able to provide here.

Generic data structures were also modified to take advantage of properties inherent to 
a DNA \kmer set, 
either simply that the strings are of fixed length or, more strongly, have the \subprop.
The most notable examples of this were the works by~\citet{pellow2016} to modify Bloom filters, by~\citet{holley2016bloom} to modify string tries (i.e. the Bloom filter trie data structure),
and by~\citet{BOSS2012} to modify the FM-index (i.e. BOSS data structure).
\citet{pellow2016} improved the space usage of Bloom filters, though the theoretical analysis is beyond the scope of this survey. 
The improvements of \citet{holley2016bloom} to a string trie were more practical and difficult to theoretically analyze.
\citet{BOSS2012} were able to simultaneously achieve the $\bigoh(n)$ space usage of Bloom filters and the perfect accuracy of a hash table, 
without affecting the query times. 
This, however, does not hold in the worst case because it assumes that the 
number of sources and sinks in the de Bruijn graph is negligible.
Later papers showed how to modify BOSS to achieve different trade-offs~\citep{li2016megahit,biBOSS}.

There were also two novel types of data structures developed specifically for the \kmer setting.
The first was unitig-based representations, proposed by~\citet{CLJSM2014} and 
later extended to spectrum-preserving string set representations by~\citep{brindathesis,spss,simplitigs}.
These representations work by first constructing the unitigs and then building an index on top of them.
The type of index varies: the FM-index is used by~\citet{CLJSM2014} and~\citet{guo18}, while
a minimum perfect hash function is used by~\citet{putterfish},~\citet{blight}, and~\citet{bifrost}.
Unitig-based representations were specifically designed to exploit the \subprop to save space, resulting in $\bigoh(n + U(k-1))$ space
($U$ is the number of maximal unitigs in the input).
Membership and navigation remain efficient ($\bigoh(k)$ and $\bigoh(1)$, respectively), 
except that for \kmers at the boundaries of unitigs, navigation takes $\bigoh(k)$.
The idea is that in practice, the \subprop implies that $U$ is much smaller than $n$,
resulting in low space and making boundary \kmers rare in practice.
A direct comparison between unitig-based representations and other representations (e.g. BOSS) to determine the regimes in which one outperforms the other has not,
to the best of our knowledge,  been attempted;
this includes either a theoretical or a comprehensive empirical analysis.

The second type of data structure developed specifically for a \kmer set is a navigational data structure,
which exploits the way that a DNA \kmer set is often queried.
These data structures retain $\bigoh(1)$ navigational queries but sacrifice the efficiency and/or feasibility of membership queries in order to achieve $\bigoh(n)$ space.
\citet{minia} were the first to use such a data structure, and~\citet{CLJSM2014} later formalized the idea; 
other navigational data structures were later developed by
~\citet{spades},
~\citet{cascading}, 
~\citet{jackman2016abyss},
~\citet{fdbg}, and
~\citet{LCBRP2016}.

Reading through the literature in this field, 
one often encounters papers on the representation of de Bruijn graphs as opposed to representation of a \kmer set.
The distinction between the two is unclear to us, as a de Bruijn graph and a \kmer set represent equivalent information (i.e. there is a bijection between the universe of \kmer sets and the universe of de Bruijn graphs).
One distinction may be that the term ``de Bruijn graph'' implies that edge queries (which in the node-centric version correspond to navigational queries, in our terminology) are efficient, 
while the term ``\kmer set'' does not connote anything about navigation.
On the other hand, ``de Bruijn graph'' obfuscates the fact that there are no degrees of freedom in defining the edge set:
once the node labels (i.e. \kmers) are determined, so are the edges.
This is in the node-centric setting, but in the edge-centric setting, 
it is the nodes that are determined once the edge labels (i.e. \kmers) are fixed.

Beyond the data structures, we also discussed what is known about space lower bounds.
Unfortunately, there have been only limited results. 
Besides the basic information-theoretic lower bound by~\citet{CB11}, nothing is known for membership data structures. 
For navigational data structures,~\citet{CLJSM2014} provided some lower bounds; 
however, these are of limited practical use because they only consider worst-case lower bounds, which are easily beat on real data.
Within the confines of spectrum-preserving string set representations, 
instance specific lower bounds were successfully applied empirically demonstrate 
the near-optimality of the greedy representation, on real data.

In this survey, we did not discuss in any detail how DNA \kmer sets are used in practice; 
we assume that there is some algorithm that takes a set of reads and extracts a 
\kmer set from them in a way that is useful to downstream algorithms. 
However, bringing such algorithms into some kind of unified framework would be a fascinating topic for another survey.

We hope that this area receives more systematic attention in the future, 
as \kmer set representations underly many bioinformatics tools. 
This might include expanding the set of operations beyond what we have described here, 
to better capture the way a DNA \kmer set is used.
Another promising avenue of research is to better and more explicitly model the distribution of 
\kmer sets that arise in sequencing data; 
such models can then uncover more efficient representations as well as provide useful lower bounds.
Progress in the field can also come through the creation of benchmarking datasets and 
through impartial competitive assessment of existing tools (e.g. as in~\citet{cami,assemblathon2}).
The ultimate goal though remains practical: 
to come up with data structures that improve space and query time of existing ones.

%%%%%%%%%%%%%%%%%%%%%%%%%%%%%%%%%%%%%%%%%%%%%%%%%%%%%%%%%%%%%%%%%%%%%%%%%%%%%%%%
\bibliographystyle{ACM-Reference-Format}
%\citestyle{acmnumeric}

%\bibliographystyle{alpha}
%\bibliographystyle{apalike}
\bibliography{dBGsurvey}

%%%%%%%%%%%%%%%%%%%%%%%%%%%%%%%%%%%%%%%%%%%%%%%%%%%%%%%%%%%%%%%%%%%%%%%%%%%%%%%%
%%%%%%%%%%%%%%%%%%%%%%%%%%%%%%%%%%%%%%%%%%%%%%%%%%%%%%%%%%%%%%%%%%%%%%%%%%%%%%%%
%%%%%%%%%%%%%%%%%%%%%%%%%%%%%%%%%%%%%%%%%%%%%%%%%%%%%%%%%%%%%%%%%%%%%%%%%%%%%%%%
%%%%%%%%%%%%%%%%%%%%%%%%%%%%%%%%%%%%%%%%%%%%%%%%%%%%%%%%%%%%%%%%%%%%%%%%%%%%%%%%
%%%%%%%%%%%%%%%%%%%%%%%%%%%%%%%%%%%%%%%%%%%%%%%%%%%%%%%%%%%%%%%%%%%%%%%%%%%%%%%%
\clearpage
\appendix

%%%%%%%%%%%%%%%%%%%%%%%%%%%%%%%%%%%%%%%%%%%%%%%%%%%%%%%%%%%%%%%%%%%%%%%%%%%%%%%%

\section{Derivations of complexities}

%%%%%%%%%%%%%%%%%%%%%%%%%%%%%%%%%%%%%%%%%%%%
\subsection{Conway and Bromage}
\citet{CB11} present separate structures for dense and sparse sets;
in our case, the sparse bitmap representation (called sarray in \citet{CB11}) is relevant.
The space taken by sarray is given in Table 1 of \citet{CB11} as
$
\mu \log \frac{\nu}{\mu}+1.92\mu + o(\mu) 
$.
In our case, $\mu = n$ and $\nu = \sigma^k$.
Membership is implemented as a constant number of rank operations, which are supported in sarray in time
${\mathcal O}(\log \frac{\nu}{\mu})+{\mathcal O}(\log^4\mu / \log \nu)$
(Table 1 in~\citet{CB11}).
In terms of construction time, we did not find an analysis in either \citet{CB11} or \citet{okanohara2007practical}.
We show the construction time as $\Omega(nk)$, since it is at least necessary to hash each \kmer. 

\subsection{Bloom filter tries}
The Bloom filter trie complexities depend on several internal parameters (e.g. $\ell,c,f,q,\lambda$ in the paper).
For our analysis, we have treated these as constants, and,  
in particular, we have set $\ell=1$ as it minimizes the complexity of operations. 
Yet, this is an extreme case that has not been explicitly considered in the original article, and
\citet{holley2016bloom} suggested optimizations for performing faster navigational queries
that are not reflected by our analysis here.
A more fine-grained analysis then we have done here is likely possible, in terms of these internal parameters.

\subsection{BOSS}

In \citet{BOSS2012}, the time complexity of $\membership(x)$ query (called $\textit{Index}(x)$) is ${\mathcal O}(k(t_f+t_b(m,2\sigma))$, where $t_f$ is ${\mathcal O}(1)$ (rank \&{} select \citep{rrr}) for the static case and ${\mathcal O}(\log\sigma)$ (a balanced binary search tree) for the dynamic case, and $t_b$ is the maximum of complexities of functions rank, select, and access on strings, which is ${\mathcal O}(\frac{\log \sigma}{\log\log n})$ for the static implementation \citep{FMMN2007} and ${\mathcal O}(\frac{\log n}{\log\log n}(1+\frac{\log\sigma}{\log\log n}))$ for the dynamic implementation \citep{NS2014}. Considering that the alphabet size is constant in our case, the static implementation makes $\membership(x)$ query time complexity equal to ${\mathcal O}(k)$ and the dynamic complexity makes it ${\mathcal O}(k(1+\frac{\log n}{\log\log n}))$.

The time complexity of $\forward(x,a)$ query (called $\textit{Outgoing}(x,a)$) is ${\mathcal O}(t_f+t_b(m,2\sigma))$, 
which is ${\mathcal O}(1)$ for the static case and ${\mathcal O}(\frac{\log n}{\log\log n})$ for the dynamic case.
The time complexity of $\backward(x,a)$ query (called $\textit{Incoming}(x,a)$) is ${\mathcal O}(k(t_f+t_b(m,2\sigma))\log\sigma)$, 
which is ${\mathcal O}(k\log\sigma)$ for the static case and ${\mathcal O}(k\log\sigma(1+\frac{\log n}{\log\log n}))$ for the dynamic case.
Both static \citep{FMMN2007} and dynamic \citep{NS2014} rank \&{} select implementations have the same asymptotic space complexity; therefore, both the static and dynamic BOSS have the same asymptotic space complexity.

\subsection{variable-order BOSS}
\sloppypar
In the case of a constant alphabet, the variable-order BOSS \citep{BBGPS2015} representation uses the data structures of original BOSS and a new $L^*$ array requiring ${\mathcal O}(n\log K)$ space \citep[Theorem 1]{BBGPS2015}. The $\membership(x)$ query is used in the same way as in BOSS. Operations $\forward(x,a)$ and $\backward(x,a)$ for $K$-mers are also used in the same way as in BOSS. For $k$-mers with $k<K$ the operations are implemented in a different (slower) way: $\forward(x,a)=\shorter(\forward(\maxlen(x,a),a),k_v)$, $\backward(x)=\shorter(\backward(\maxlen(\longer(x,k_v+1),*)),k_v)$, $\lastchar(x)=\lastchar(\maxlen(x,*))$. Note, $\backward(x)$ in the variable-order BOSS returns a list of nodes with an edge to $x$. In \citep[Section~5]{BBGPS2015} variable-order BOSS $\backward(x)$ time complexity is ${\mathcal O}(\sigma(t_{\backward(x)}+\log k))$. Operation $\maxlen([i,j],a)$ runs in ${\mathcal O}(\log |\Sigma|)$ time (i.e. ${\mathcal O}(1)$ time for $|\Sigma|=\text{const}$), $\maxlen([i,j],*)$ runs in ${\mathcal O}(1)$ time. 
Operation $\shorter([i,j],k)$ runs in time ${\mathcal O}(\log K)$ and operation $\longer([i,j],k)$ runs in time ${\mathcal O}(|B|\log K)$, where $B$ is a range of nodes sharing the same suffix of length $k$.
\newpage

\end{document}